\documentclass[12pt,english,aps,twocolumn, 3p]{revtex4}
\usepackage[T1]{fontenc}
\usepackage[latin9]{inputenc}
\usepackage[a4paper]{geometry}
\usepackage{babel}
\usepackage{float}
\usepackage{textcomp}
\usepackage{amstext}
\usepackage{graphicx}
\usepackage{esint}
\usepackage[unicode=true,
 bookmarks=true,bookmarksnumbered=false,bookmarksopen=false,
 breaklinks=false,pdfborder={0 0 1},backref=false,colorlinks=false]
 {hyperref}

\makeatletter
\@ifundefined{textcolor}{}
{%
 \definecolor{BLACK}{gray}{0}
 \definecolor{WHITE}{gray}{1}
 \definecolor{RED}{rgb}{1,0,0}
 \definecolor{GREEN}{rgb}{0,1,0}
 \definecolor{BLUE}{rgb}{0,0,1}
 \definecolor{CYAN}{cmyk}{1,0,0,0}
 \definecolor{MAGENTA}{cmyk}{0,1,0,0}
 \definecolor{YELLOW}{cmyk}{0,0,1,0}
}

\makeatother

\begin{document}

\title{Nanometer-scale Tomographic Reconstruction of 3D Electrostatic Potentials
in GaAs/AlGaAs Core-Shell Nanowires}

\author{A. Lubk}

\affiliation{Triebenberg Laboratory, Institute of Structure Physics, Technische
Universität Dresden, 01062 Dresden, Germany}

\author{D. Wolf}

\affiliation{Triebenberg Laboratory, Institute of Structure Physics, Technische
Universität Dresden, 01062 Dresden, Germany}

\author{P. Prete}

\affiliation{IMM-CNR, Lecce Research Unit, S.P. 6 Lecce-Monteroni, I-73100, Lecce,
Italy}

\author{N. Lovergine}

\affiliation{Dipartimento di Ingegneria dell\textquoteright{}Innovazione, Università
del Salento, S.P. 6 Lecce-Monteroni, I-73100 Lecce, Italy}

\author{T. Niermann}

\affiliation{Institut für Optik und Atomare Physik, Technische Universität Berlin,
Straße des 17. Juni 135, 10623 Berlin, Germany }

\author{S. Sturm}

\affiliation{Triebenberg Laboratory, Institute of Structure Physics, Technische
Universität Dresden, 01062 Dresden, Germany}

\author{H. Lichte}

\affiliation{Triebenberg Laboratory, Institute of Structure Physics, Technische
Universität Dresden, 01062 Dresden, Germany}
\begin{abstract}
\textbf{We report on the development of Electron Holographic Tomography
towards a versatile potential measurement technique, overcoming several
limitations, such as a limited tilt range, previously hampering a
reproducible and accurate electrostatic potential reconstruction in
three dimensions. Most notably, tomographic reconstruction is performed
on optimally sampled polar grids taking into account symmetry and
other spatial constraints of the nanostructure. Furthermore, holographic
tilt series acquisition and alignment have been automated and adapted
to three dimensions. We demonstrate $6$ nm spatial and $0.2$ V signal
resolution by reconstructing various, previously hidden, potential
details of a GaAs/AlGaAs core-shell nanowire. The improved tomographic
reconstruction opens pathways towards the detection of minute potentials
in nanostructures and an increase in speed and accuracy in related
techniques such as X-ray tomography. }
\end{abstract}
\maketitle
Tomographic techniques use lower dimensional projections of $n$-dimensional
data to reconstruct the original quantity. This principle dates back
to the work of the Austrian mathematician J. Radon \cite{Radon(1917)}
and to date applications in fields as diverse as medicine, geophysics,
material science and quantum information are reported. Our main goal
is the tomographic reconstruction of 3D electrostatic potentials in
nanostructures. They are tightly connected to the chemical composition
and electronic structure and therefore mirror the corresponding functionality
and possible failures. Off-axis electron holography (EH) provides
unique access to these fields \cite{Dunin-Borkowski_EoNaN3(2004)41,Lichte_RoPiP71(2008)16102,Midgley_NM8(2009)271}
because the reconstructed phase is in the phase grating approximation
proportional to the potential projected along lines $l$ 
\begin{equation}
\varphi\left(x,p\right)=\left(C_{E}\int_{l\left(x,p,\theta\right)}V\left(\mathbf{r}\right)\mathrm{d}s\right)\,\mathrm{mod}\left(2\pi\right)\label{eq:phase}
\end{equation}
which holds for a wide range of imaging conditions \cite{Matteucci(2002)}.
Here, $C_{E}$ denotes the electron-matter interaction constant depending
only on the acceleration voltage of the electrons ($C_{E}=6.5$~mrad/(Vnm)
@ 300 kV), $\left(x,p,\theta\right)^{T}$ the coordinates of the detector,
and $\mathbf{r}=\left(x,y,z\right)^{T}$ the coordinates in real space
as explained in Fig. \ref{fig:Holographic-tomographic-principl}.
In particular the $x$-axis is set parallel to the tilt axis in order
to reduce the 3D reconstruction to a slice by slice 2D tomographic
reconstruction problem in planes perpendicular to the tilt axis. 

Following the mathematical foundations of tomography (e.g. \cite{Natterer(2001)}),
the collection of projected potentials (\ref{eq:phase}) under different
angles $\theta$ represents the Radon transform $\hat{V}=\mathit{\mathcal{R}}\left\{ V\right\} $
of $V$. The inversion of this transformation (inverse Radon transformation)
then yields the potential $V=\mathit{\mathcal{R}}^{-1}\left\{ \hat{V}\right\} $
forming the basis of EH tomography (EHT). The proof-of-concept for
EHT dates back to the work of Lai et al. \cite{Lai(1994)a}, however,
quantitative reconstructions \cite{Twitchett-Harrison_NL7(2007)2020,Fujita_JoEM58(2009)301,Wolf_UM110(2010)390}
were delayed for a long time mainly for two reasons: first, the instrumental
demands, such as specific tomography specimen holders allowing ultrahigh
tilt angles, nanometer precise computer-controlled goniometers as
well as powerful computers had to be developed. Second, the EHT procedure
is very comprehensive and time consuming.

The most notable experimental limitation of virtual all electron-microscopic
tomographic techniques is the limited tilt range of usually \textpm{}70\textdegree{}
instead of \textpm{}90\textdegree{}. This leads to a loss of information,
visible as \textquoteleft{}\textquoteleft{}missing wedge\textquoteright{}\textquoteright{}
in the Fourier transform of the tomogram (see e.g. \cite{Midgley_UM96(2003)413}).
In real space, this corresponds to a reduced resolution in the tomogram
in the direction of the missing wedge. Moreover, dynamical scattering
effects, tilt axis misalignment, phase unwrapping failures and shot
noise introduce errors in the projected data that can amplify upon
reconstruction because of the (mildly) ill-conditioned nature of the
Radon transform \cite{Natterer(2001)}. Similar to other tomographic
techniques, large and ongoing efforts are therefore put into developing
regularization schemes ($\cong$suppressing the error amplification)
typically involving auxiliary conditions such as (euclidean) norm
minimization ($\cong$Tikhonov regularization \cite{Tikhonov(1977)}),
total variation minimization ($\cong$generalized Tikhonov regularization
\cite{Goris(2012)}) or signal range restriction (discrete tomography
\cite{Batenburg(2009)}). Since such conditions typically introduce
an additional regularization error, the art of regularization consists
of minimizing the total error of both original and regularization
error. That task is further complicated by the fact that these errors
are generally unknown \cite{Hansen(1998)}. It is therefore always
advantageous to use preferably error-free conditions based on real
physical properties of the potential such as physical spatial constraints
(e.g. \cite{Twitchett-Harrison_NL7(2007)2020}). That can be due to
outer or internal boundaries of nanoparticles or symmetries imposed
by the crystal structure. 

How spatial constraints can be used to improve the tomographic reconstruction
will be discussed in the following section. We will furthermore emphasize
the importance of sampling, that is, the relation between detector
resolution, number of projection angles and reconstructed object features,
for the design of optimal reconstruction algorithms. The obtained
improvement will be valuable for all types of tomography in spite
of our particular application to EHT. Our second focus are improving
methods for phase unwrapping, dynamic scattering correction and tilt
series alignment, mainly dedicated to EHT. The combination of which
will significantly improve EHT as 3D potential reconstruction method.

Using the optimized reconstruction procedure we successfully retrieved
the 3D potential of a $\left\langle 111\right\rangle $-oriented GaAs-Al$_{0.33}$Ga$_{0.67}$As
core-shell nanowire (NW) grown by metal-organic vapor phase epitaxy
(MOVPE) through a Au nanoparticles (NP) as metal catalyst (see Ref.
\cite{Prete_JoCG310(2008)5114,Wolf(2011)}). The structure provides
a wide range of electrostatic potentials with sharp interfacesas well
as long rang gradients. The AlGaAs shell contains also self-assembled
Al segregations and local alloy fluctuations due to the different
Ga and Al adatom mobilities on the nanowire surface \cite{Rudolph(2013),Heiss_NM12(2013)439}.
Note that these features are crucial for envisaged applications of
semiconductor NW in future nano-scaled electronic, optoelectronic,
and photovoltaic devices \cite{Bryllert_NT17(2006)227,Gallo_APL98(2011)241113,Krogstrup_NP7(2013)306}.
Moreover, we previously investigated this system with less developed
EHT techniques \cite{Wolf(2011)}, rendering it the ideal test case
for verifying the achieved progress.

\section{Algebraic Holographic Tomography}

\begin{figure}[H]
\includegraphics[bb=14bp 60bp 492bp 320bp,clip,scale=0.5]{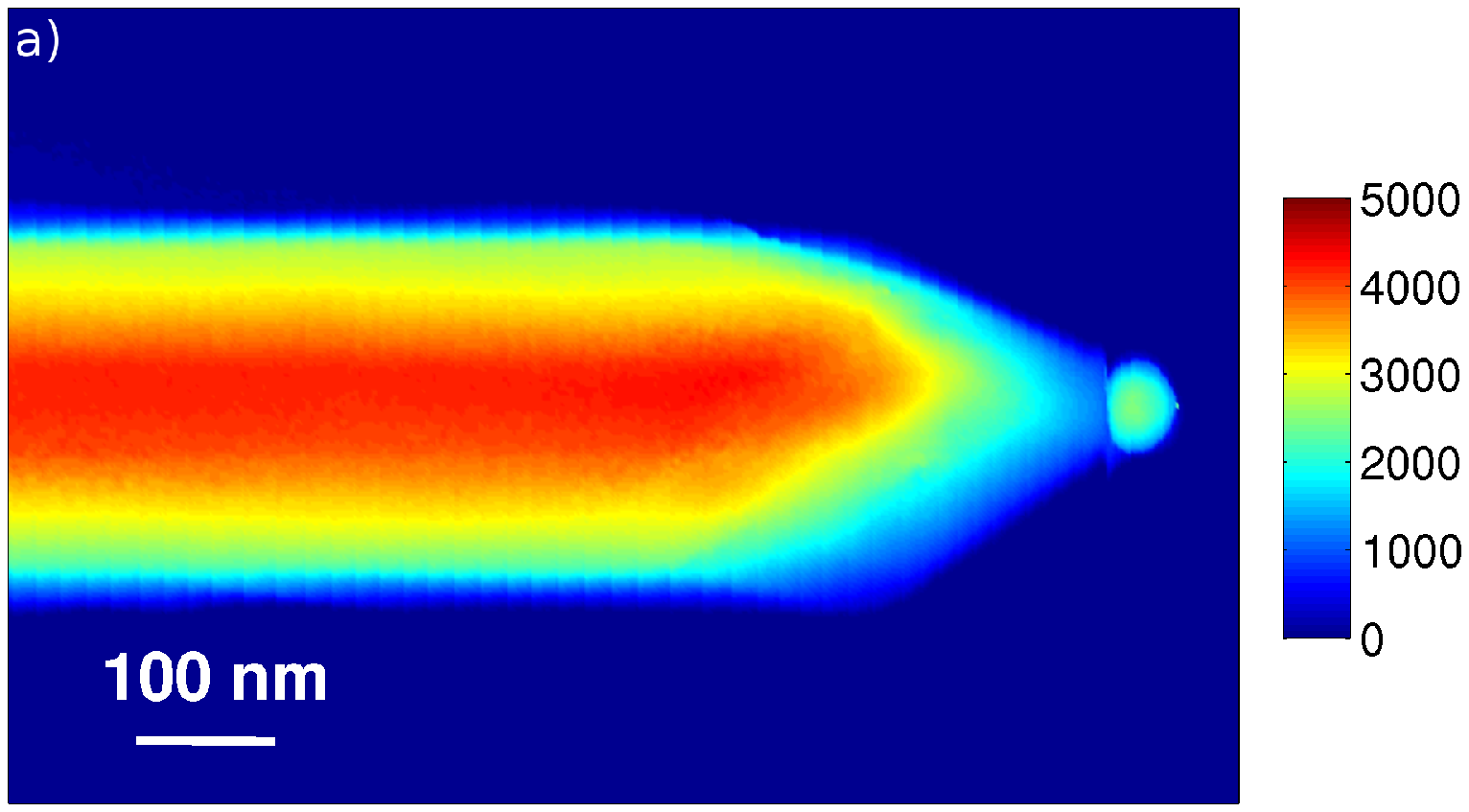}

\includegraphics[bb=14bp 60bp 492bp 320bp,clip,scale=0.5]{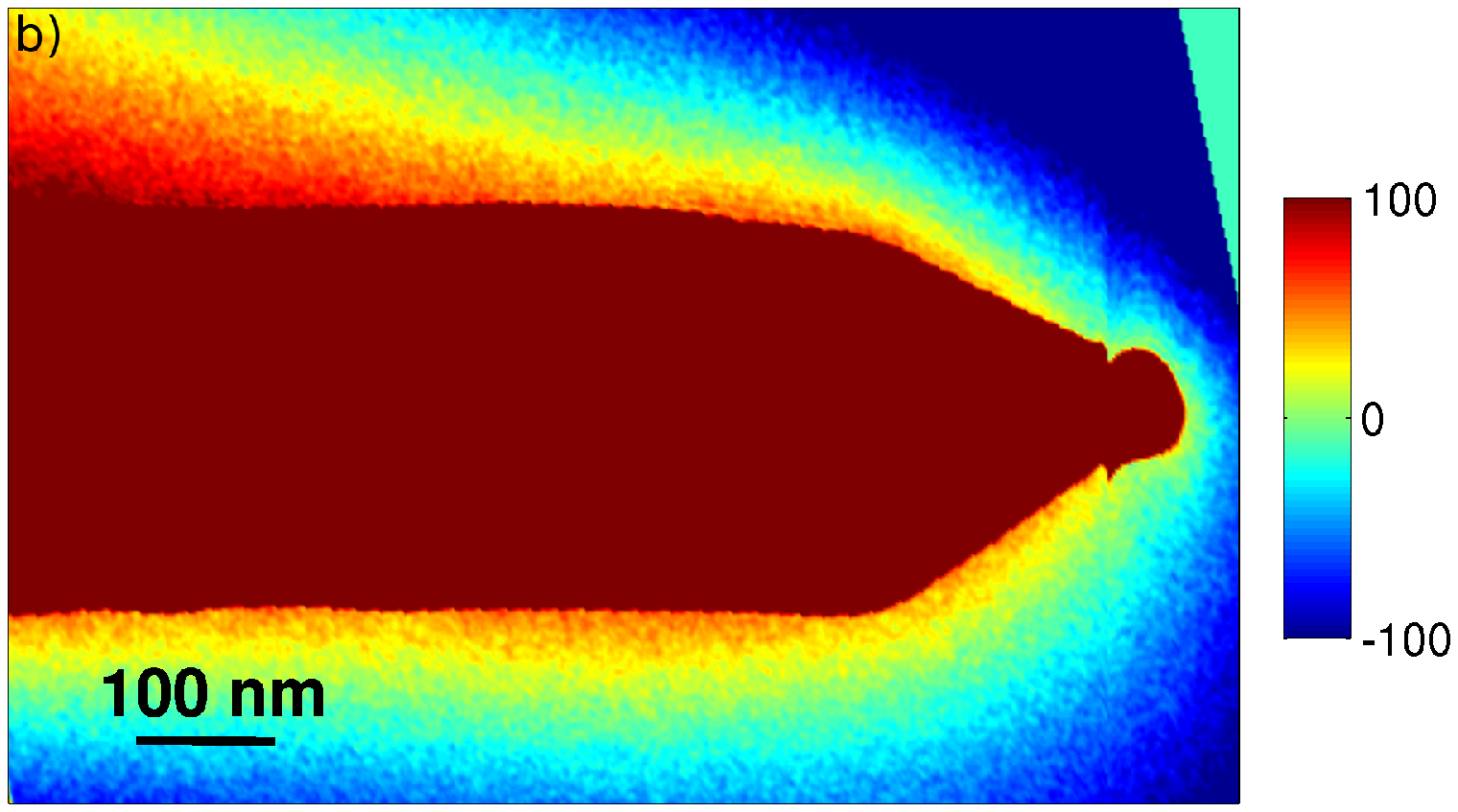}\caption{\label{fig:Examplary-projected-potential}Examplary projected potential
(in Vnm) reconstructed from a hologram of the tilt series (at 30\textdegree{}
tilt angle) depicted at two different colour scales to illustrate
the magnitude of the charging. On top the complete potential reachig
several kVnm in the NW is depicted. The charging field below is approximately
2 order of magnitude lower and is also slightly superimposed by an
additional phase wedge from the imperfect alignment of the sideband.
The zero potential region on the top right stems from the alignment
of the tilt axis along $x$.}
\end{figure}

\begin{figure*}[t]
\includegraphics[bb=30bp 530bp 565bp 810bp,clip,scale=0.9]{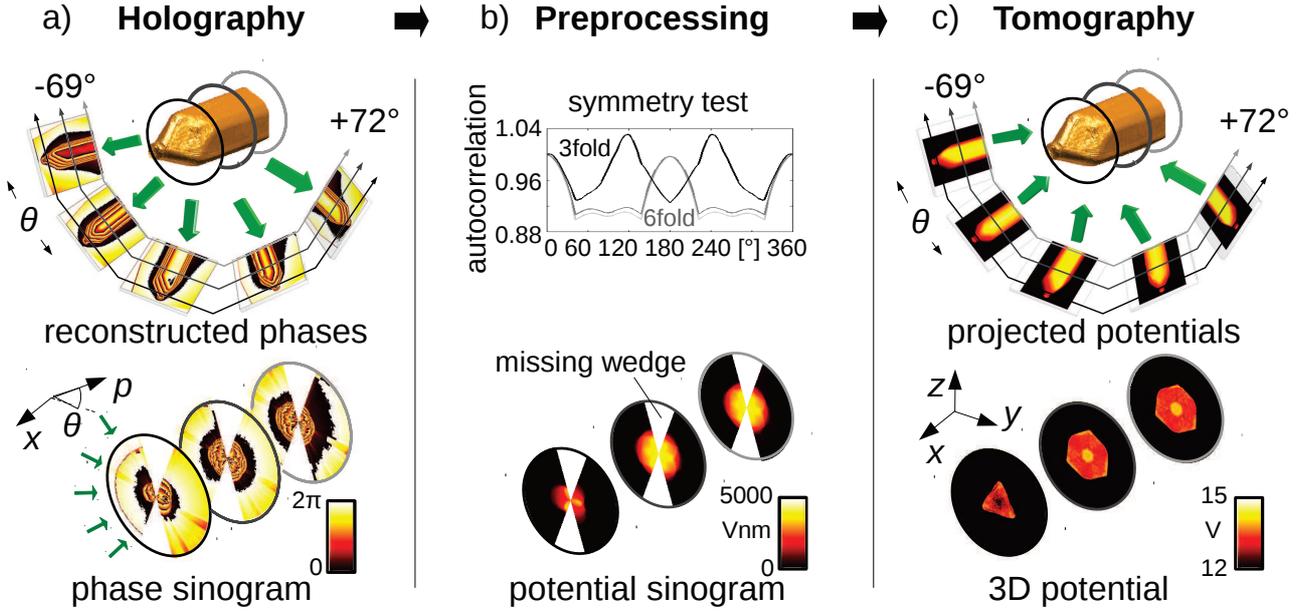}\caption{Holographic tomographic principle: a) Holographic tilt series (top)
and selected sinograms (bottom). b) Rotational symmetry test (top)
on preprocessed projected potential data (bottom). The 6-fold (hexagonal)
symmetry of the NW is partly suppressed by the 2-fold symmetry of
the missing wedge. c) Tomographic reconstruction (top) and reconstructed
potential (bottom). Throughout the figure the 5 green arrows denote
a set of angles ($-66\text{\textdegree},\,-33\text{\textdegree},\,0\text{\textdegree},\,33\text{\textdegree},\,66\text{\textdegree}$)
and 3 deliberately chosen slices at $x=100,\,400,\,700$ nm are highlighted
by black, dark gray and light gray lines and frames. \label{fig:Holographic-tomographic-principl}}
\end{figure*}

The general workflow of EHT consisting of hologram tilt series acquisition,
phase reconstruction from hologram tilt series; phase tilt series
alignment and tomographic reconstruction is depicted in Fig. \ref{fig:Holographic-tomographic-principl}
using the example of the GaAs-AlGaAs NW. All steps have been widely
automated, not only to decrease the amount of time from initially
$\mathcal{O}\left(10\,\mathrm{h}\right)$ to presently $\mathcal{O}\left(1\,\mathrm{h}\right)$
(see \cite{Wolf_UM110(2010)390} for details), but also to increase
the reproducibility and quality of the obtained data. Thus, the automated
workflow is an important prerequisite towards widespread application.
The holographic tilt series of an individual NW with an entire diameter
of 300~nm (80~nm core, 110 nm shell) was recorded in a range from
$-69\text{\textdegree}$ to $+72\text{\textdegree\ in \ensuremath{3\text{\textdegree}}}$
steps at the FEI Titan 80-300 Berlin Holography Special TEM using
the THOMAS software package \cite{Wolf_UM110(2010)390}. The TEM was
operated at 300 kV acceleration voltage in aberration corrected Lorentz
mode that provides a resolution of about $2\,\mathrm{nm}$. The holograms
were acquired with a double biprism setup with a field of view of
1 $\mathrm{\mu}$m and a fringe spacing of 3 nm. The object exit wave
tilt series has been reconstructed from the object holograms and corresponding
empty holograms applying a Butterworth filter of $0.13\,\mathrm{nm^{-1}}$
FWHM to separate the side band (see e.g. in \cite{Lehmann_MaM8(2002)447}
for further details of the holographic method). This was performed
automatically for the entire tilt series. Fig. \ref{fig:Examplary-projected-potential}
shows a typical example of an holographically reconstructed projected
potential of the NW. To highlight the effects in the vacuum region
due to beam charging and remaining phase wedges due to the sideband
alignment we also show the outer region in Fig. \ref{fig:Examplary-projected-potential}.
Here, one observes a small charging field accumulating around 50 Vnm
at the edge of the NW and dropping quickly to -50 Vnm towards the
rim of the reconstruction volume. Considering that the whole NW has
a diameter of 300 nm this corresponds to a surface potential due to
charging of around 0.3 V. To test its influence on the tomographic
reconstruction we performed the latter with and without constraining
the complete vacuum to 0 V as stated in the results section below. 

Following holographic wave reconstruction and reorientation of the
tilt axis along the $x$-axis, three important processing steps determine
the accuracy of the subsequent tomographic reconstruction: (1) The
phase needs to be unwrapped to obtain the projected potential from
the reconstructed waves. The unwrapping is problematic if the sampled
phase difference between two pixels exceeds $\pi$ \cite{Ghiglia(1998)},
a situation often present in noisy data with sharp thickness jumps
at specimen edges. We tackled this problem by extending the original
2D unwrapping performed at each holographic reconstruction to 3D with
the tilt angle as third dimension (i.e. the whole tilt series). That
is possible because of the continuous dependency of the phase on the
tilt angle and yielded accurate projected potentials even at the Au
catalyst. The next step (2) consists of suppressing dynamical scattering
artifacts producing oscillations of the phase close to zone axis conditions
(Fig. \ref{fig:Holographic-tomographic-principl}a)). The latter could
be largely removed by normalizing the average of the projected potentials
at all angles to the same value, in agreement with a dynamical correction
factor approach \cite{Lubk(2010)}. (3) Finally, this data is aligned
around a common tilt axis with the center of mass method. This alignment
was performed slice by slice for sampling reasons discussed below.
The tilt series ready for tomographic reconstruction is depicted in
Fig. \ref{fig:Holographic-tomographic-principl}b. Besides the substantial
improvement of the data one observes a 6-fold symmetry in the core-shell
region reducing to a 3-fold symmetry in the tapered section below
the Au NP (corroborated by azimuthal cross-correlation in Fig. \ref{fig:Holographic-tomographic-principl}b).

Now the actual tomographic reconstruction starts. The variety of reconstruction
techniques can be separated into filtered, Fourier and algebraic algorithms,
which coincide in the limit of infinitely fine sampling. Fourier methods
and filtered techniques, such as Weighted Back Projection (WBP) \cite{Gilbert_PRSL.182(1972)89,Radermacher(2006)}
or Weighted Simultaneous Reconstruction Techniques (W-SIRT) \cite{Wolf(2014)}
are based on numerically efficient and fast adaptions of analytic
Radon transformation formulas. Algebraic reconstructions consider
the sampled Radon transformation as a large system of linear equations
which are typically inverted by iterative algebraic techniques. Here,
we will implement and further develop the latter because of its higher
accuracy and flexibility at the cost of only minor reduction in speed.

\begin{figure*}[t]
\includegraphics[bb=30bp 230bp 810bp 560bp,clip,width=0.8\paperwidth]{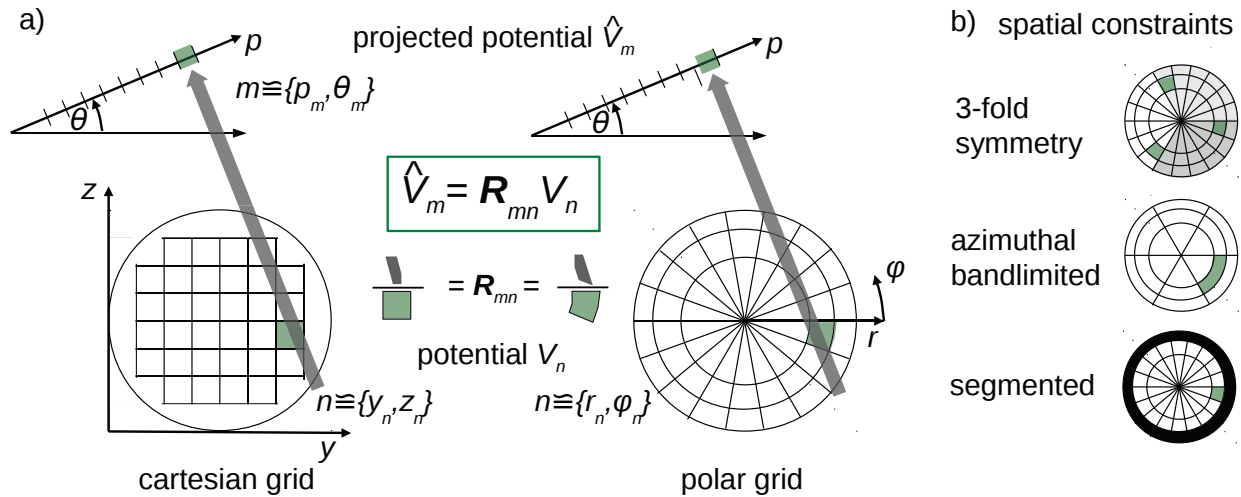}\caption{Algebraic tomography and sampling: a) A deliberately chosen sampling
(left: Cartesian, right: polar) with sampling index $n$ is projected
under a certain angle $\theta_{m}$ on the detector pixel $p_{m}$
with the projection weight of the reconstructed pixel $n$ determined
by the normalized covering of the projection ray. b) Three examples
of spatial constraints applied to the polar sampling scheme from top
to bottom: 3-fold symmetric projection of 3 symmetric pixels into
corresponding detector pixels; azimuthal band-limited projection facilitating
reduction of azimuthal sampling points; deliberate exclusion of potential
free region (e.g. vacuum) from Radon matrix. \label{fig:algebraic-tomography}}
\end{figure*}

In Fig. \ref{fig:algebraic-tomography} the main idea behind algebraic
reconstruction is illustrated: Here, the set of all sampled projections
with index $m$ is the algebraic version of the Radon transform $\hat{V}_{m}=\mathcal{R}_{mn}V_{n}$
of the deliberately sampled potential (index $n$). The Radon matrix
is usually sparse and tomographic reconstruction corresponds to finding
a (generalized) inverse to the latter. Although sampling the position
space by a the Cartesian grid is by far dominant in applications we
will show in the following that the polar grid is the better choice
indeed: 

(i) First of all there is a deep connection between the polar grid
and sampling theorems valid for 2D Radon transformations. The latter
state that tomographic reconstruction from finite tilt angles requires
azimuthally band-limited data \cite{Natterer(2001)} (see also Supplementary
Information). From that it follows that the axis with the smallest
azimuthal band limit not only determines the number of required tilt
angles in the experiment but also the number of azimuthal grid points
in the reconstruction on a polar grid. For example, a radially symmetric
object (e.g. a homogeneous cylinder) aligned along the symmetry axis
can be reconstructed from a single projection only \cite{Grillo(2013)}.
Accordingly, some large annular pixels with radial dimensions determined
by the radial resolution are sufficient when reconstructing this function
on a polar grid centered around the symmetry axis. Obviously the number
of Cartesian pixels would have been much larger in this case. By aligning
the tilt series slice-wise around the center of mass we closely approximate
the 3-fold symmetry axis of the NW facilitating a beneficial use of
the polar grid in terms of minimizing pixel numbers in position space.
By this, memory requirements and computing time for matrix inversion
(typically scaling with (matrix dimension)$^{2\ldots3}$) could be
greatly reduced. (ii) It is immediately obvious that the above noted
spatial constraints can be straight forwardly implemented by shaping
the projected and reconstructed pixels correspondingly. Fig. \ref{fig:algebraic-tomography}b)
illustrates sampling schemes for $n$-fold rotationally symmetric
objects, azimuthal band limited objects and segmented objects. Indeed,
one can deliberately combine different constraints. Beside these two
main advantages we note that the radial grid exploits the whole circular
reconstruction area.
 For the (generalized) matrix inversion itself a large number of numerical
methods has been developed. For tomography the Kaczmarz ($\cong$Algebraic
Reconstruction Technique - ART) and related methods (e.g. Sequential
Iterative Reconstruction Technique - SIRT) are very popular \cite{Natterer(2001)},
probably due to its particularly easy implementation and built-in
quasi Tikhonov regularization parametrized by the number of iterations
\cite{Hansen(1998)}. The latter refers to a a growing reconstruction
of larger spatial frequency components at higher iteration numbers,
implying a growth of spatial resolution at cost of increasing noise.
However, the Kaczmarz method is not seeking for the optimal (fastest)
iteration and is not optimized for sparse matrices such as the Radon
matrix. Therefore, we use a conjugate gradient method implemented
in the LSQR algorithm \cite{Paige(1982)} to reduce significantly
the number of iterations (due to a weaker quasi Tikhonov regularization)
and thus computing time.

\section{Results}

\begin{figure*}[t]
\includegraphics[bb=40bp 540bp 550bp 810bp,clip,scale=0.9]{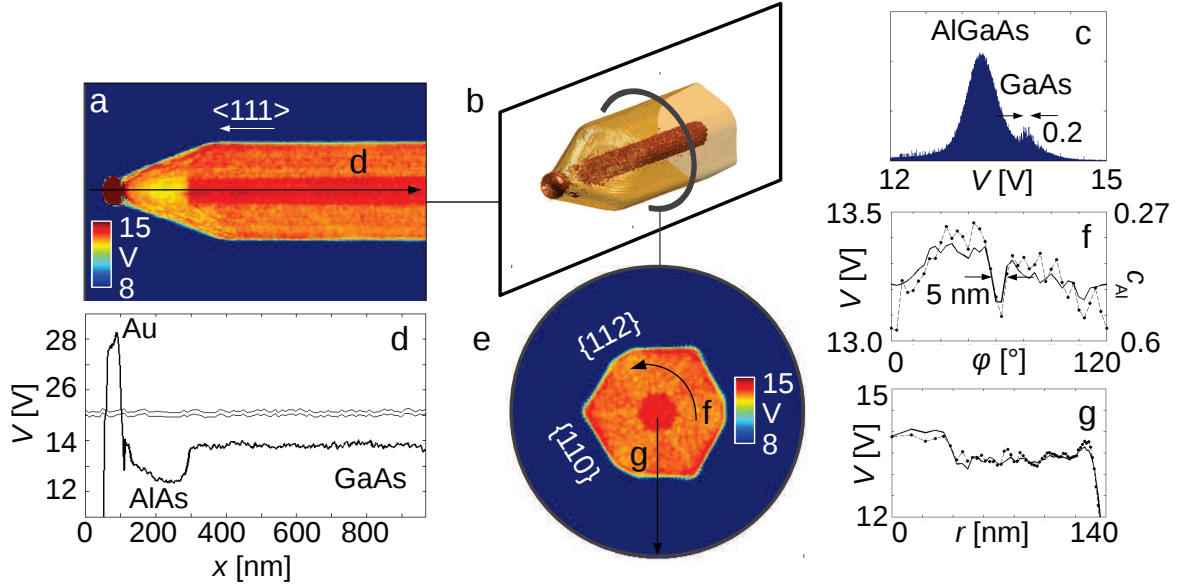}\caption{\label{fig:results}Reconstructed 3D potential of GaAs-Al$_{0.33}$Ga$_{0.67}$As
core shell NW: The isosurfaces (b) at 9 and 13.75 V illustrate the
core-shell morphology corroborated by the double peaks in the corresponding
histogram (c). The $zx$-plane (a) with corresponding linescan (d)
on the left shows the sequence Au, AlAs and GaAs unraveled through
their different MIPs. The projected $yz$-plane (e) on the right with
corresponding linescans in azimuthal (f) and radial (g) direction
shows the core-shell structure with local Al accumulation identified
through characteristic potential reduction in the azimuthal scan.
Here, the corresponding Al concentration $c_{Al}$ determined from
Vegard's Law is shown on the right y-scale. The thick lines in (f)
and (g) indicate averaged values from the whole hexagonal region of
the NW.}
\end{figure*}

We present results from a tomographic reconstruction incorporating
a 3-fold symmetry constraint and 10 LSQR iterations on a $13800\times11700$
Radon matrix for a polar grid. This choice is motivated by the good
reconstruction quality (good spatial resolution at acceptable noise)
containing all important potential features. We emphasize, however,
that the regularization strength ($\cong$the number of iterations)
can be deliberately varied; therefore we supplement reconstructions
with 5 and 15 iterations and a reconstruction without symmetry constraint
(see Supplementary Information). We furthermore note that a small
charging of the NW could be observed through stray fields in the vacuum
(see Fig. \ref{fig:Examplary-projected-potential}). Its influence
on the reconstruction was tested to be around $0.1$ V by reconstructing
with and without constraining the vacuum to $0$ V in the reconstruction.
The reconstructed 3D~potential of the GaAs-Al$_{0.33}$Ga$_{0.67}$As
NW, which is illustrated by means of isosurfaces and $yz$-slices
in Fig. \ref{fig:results}b shows a slightly deformed 6-fold symmetry
in the core-shell region corroborating the symmetry test performed
on the projection data (Fig. \ref{fig:Holographic-tomographic-principl}b).
We emphasize that due to the symmetry constraint missing wedge artifacts
are now completely absent (see \cite{Wolf(2011)} for comparison);
it would even be possible to use a trunctated tilt series interval
of $60\text{\textdegree}$ here. In total, the achieved signal precision
of about $0.2$ V (determined from the standard deviations of the
corresponding homogeneous potential region along $x$) and spatial
resolution of 6 nm (determined from the FWHM of the potential gradient
at the NW boundary) is mainly limited by the holographic imaging process
and not the EHT method; and therefore allows detecting the following
features \emph{in 3D}. 

The Au NP at the NW tip has an average potential of $27.8\pm0.2$
V (Fig. \ref{fig:results}a,d) which agrees very well with the theoretical
value of $28$ V obtained by multiplying the DFT value $30.1$ V \cite{Schowalter(2006)}
with a dynamical scattering correction factor of $0.93$ \cite{Lubk(2010)}.
The same holds for the average potential of the GaAs core ($V_{\mathrm{exp}}=13.9\pm0.1\,\mathrm{V},\, V_{\mathrm{theo}}=14.19\,\mathrm{V}$
\cite{Kruse(2006)}) and the Al$_{0.33}$Ga$_{0.67}$As shell ($V_{\mathrm{exp}}=13.4\pm0.2\,\mathrm{V},\, V_{\mathrm{theo}}=0.33\cdot12.34\,\mathrm{V}+0.67\cdot14.19\,\mathrm{V}=13.58\,\mathrm{V}$
\cite{Kruse(2006)}). %
\footnote{The small discrepancy could be due to small imperfections of the generalized-gradient
approximation of exchange-correlation in the theoretic calculations,
the choice of the phase reference in the holographic experiment or
the small charging field. %
}We furthermore observe an axial decay of the potential from the Au-Al$_{0.33}$Ga$_{0.67}$As
interface over approximately $100$ nm comprising $0.5$ V (Fig. \ref{fig:results}d),
which could indicate a Fermi-Level-pinned Metal-AlGaAs junction or
a varying chemical composition. In particular the appearance of a
small potential ($V=12.5\pm0.1\,\mathrm{V}$) region directly above
the GaAs core (Fig. \ref{fig:results}d) indicates the formation of
a AlAs alloy reported for similar GaAs-Ga$_{c}$In$_{1-c}$P core-shell
NWs \cite{Skold(2005)}. The most remarkable features are characteristic
3-fold symmetric lines of reduced potential along \{112\}-directions
of the NW (Fig. \ref{fig:results}e). This observation is in striking
agreement with the facet-dependent Al-segregation recently reported
for this system \cite{Rudolph(2013),Zheng(2013)}, and ascribed to
polarity-driven surface reconstruction during AlGaAs shell growth.
Noteworthy, in Refs. \cite{Rudolph(2013),Zheng(2013)} Al-enrichment
along \{112\}-facets was detected in the shell by means of 2D cross-sectional
STEM analysis and the determined Al concentration enhancement to $c_{Al}\approx0.67$
agree with our value $c_{Al}=0.55\pm0.5$ when taking into account
a further increase at higher iteration numbers ($\cong$ higher spatial
resolution, see Supplementary Information). 

\begin{figure}[H]
\includegraphics[bb=120bp 430bp 480bp 800bp,clip,scale=0.6]{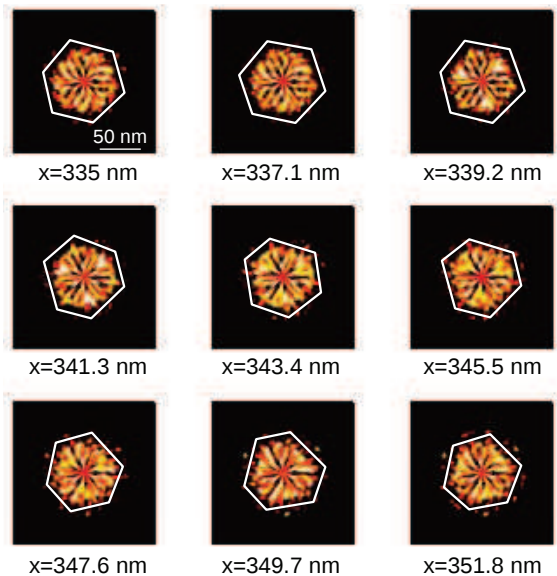}\caption{\label{fig:core}$\mathrm{GaAs}$-core showing a modulated facet structure.
White frames are drawn around the core to indicate the core facets.
The color range was restricted to an interval between 13.5 and 14.5
V to highlight the core region.}
\end{figure}

A close inspection of the 3D potential additionally revealed fluctuations
of the Al-enrichment around the above mean value. We ascribe them
to modulations of the local faceting of the core as demonstrated in
Fig. \ref{fig:core}, where we show cross-sections from the unperturbed
middle part (i.e. far away from the tapered region) of the NW. The
depicted set of cross-sections clearly reveals a changing facet structure.
Two effects can be distinguished. First there is a change in facet
length, e.g. observable in the sequence $x=335\,\mathrm{nm}$ to $x=345.5$
nm. That effect has been reported in Ref. \cite{Johansson2007} and
was there explained by alternating $\left\{ 111\right\} $-nanofacets
accumulating to the six-fold $\left\{ 110\right\} $-facets observed
on a larger length scale. Secondly, we observe a complete rotation
of about $30\text{\textdegree}$ between $x=345.5\,\mathrm{nm}$ to
$x=347.6$ possibly introduced by a twin boundary or a switch from
$\left\{ 110\right\} $-facets to $\left\{ 112\right\} $-facets.
These changes could be observed along the whole NW and are subjected
to influence the growth of the shell structure in general and the
Al-accumulation in particular.

We again emphasize that the above features have been reconstructed
in 3D for the first time (see Ref. \cite{Wolf(2011)} for comparison),
i.e. quantitative compositional changes along $x$, $y$ and $z$
are revealed in parallel without the need to destructively prepare
special cross-sections with possibly modified surfaces from the NW.

\section{Summary}

In summary we demonstrated quantitative 3D electrostatic potential
reconstruction with unprecedented accuracy and spatial resolution
by a number of separate improvements implemented into electron holographic
tomography (EHT). Most importantly, we developed automated acquisition
and wave reconstruction schemes, procedures for 3D phase unwrapping,
dynamic scattering correction, as well as optimal polar and symmetry
adapted sampling. The latter two are beneficial to all tomographic
techniques as a route towards faster and more accurate reconstructions.
For the particular case of EHT our techniques facilitate reconstruction
of nanoscale potentials such as originating from band bending or elemental
segregation where spatial and signal resolution are critical. Similarly,
3D magnetic field reconstruction can be simplified by segmentation
of the reconstruction volume into particle support and vacuum with
the latter facilitating a reconstruction without the critical subtraction
of the electrostatic phase shift.

\section{Acknowledgments}

AL, DW, and HL acknowledge financial support from the European Union
under the Seventh Framework Programme under a contract for an Integrated
Infrastructure Initiative. Reference 312483 - ESTEEM2. 8. SS is funded
by the European Union (ERDF) and the Free State of Saxony via the
ESF project 100087859 ENano. TN acknowledges financial support from
the DFG within SFB 787 and INST 131/508-1.

\bibliographystyle{apsrev}
\bibliography{manuscript}

\begin{thebibliography}{35}
\expandafter\ifx\csname natexlab\endcsname\relax\def\natexlab#1{#1}\fi
\expandafter\ifx\csname bibnamefont\endcsname\relax
  \def\bibnamefont#1{#1}\fi
\expandafter\ifx\csname bibfnamefont\endcsname\relax
  \def\bibfnamefont#1{#1}\fi
\expandafter\ifx\csname citenamefont\endcsname\relax
  \def\citenamefont#1{#1}\fi
\expandafter\ifx\csname url\endcsname\relax
  \def\url#1{\texttt{#1}}\fi
\expandafter\ifx\csname urlprefix\endcsname\relax\def\urlprefix{URL }\fi
\providecommand{\bibinfo}[2]{#2}
\providecommand{\eprint}[2][]{\url{#2}}

\bibitem[{\citenamefont{Radon}(1917)}]{Radon(1917)}
\bibinfo{author}{\bibfnamefont{J.}~\bibnamefont{Radon}}, \bibinfo{journal}{Ber.
  Verh. S\"achs. Akad. Wiss. Leipzig, Math. Nat. kl.}
  \textbf{\bibinfo{volume}{69}}, \bibinfo{pages}{262} (\bibinfo{year}{1917}).

\bibitem[{\citenamefont{Dunin-Borkowski
  et~al.}(2004)\citenamefont{Dunin-Borkowski, McCartney, and
  Smith}}]{Dunin-Borkowski_EoNaN3(2004)41}
\bibinfo{author}{\bibfnamefont{R.}~\bibnamefont{Dunin-Borkowski}},
  \bibinfo{author}{\bibfnamefont{M.}~\bibnamefont{McCartney}},
  \bibnamefont{and} \bibinfo{author}{\bibfnamefont{D.}~\bibnamefont{Smith}}, in
  \emph{\bibinfo{booktitle}{Encyclopedia of Nanoscience and Nanotechnology}},
  edited by \bibinfo{editor}{\bibfnamefont{H.}~\bibnamefont{Nalwa}}
  (\bibinfo{publisher}{American Scientific Publishers}, \bibinfo{year}{2004}),
  vol.~\bibinfo{volume}{3}, pp. \bibinfo{pages}{41--100}.

\bibitem[{\citenamefont{Lichte and Lehmann}(2008)}]{Lichte_RoPiP71(2008)16102}
\bibinfo{author}{\bibfnamefont{H.}~\bibnamefont{Lichte}} \bibnamefont{and}
  \bibinfo{author}{\bibfnamefont{M.}~\bibnamefont{Lehmann}},
  \bibinfo{journal}{Reports on Progress in Physics}
  \textbf{\bibinfo{volume}{71}}, \bibinfo{pages}{016102}
  (\bibinfo{year}{2008}).

\bibitem[{\citenamefont{Midgley and
  Dunin-Borkowski}(2009)}]{Midgley_NM8(2009)271}
\bibinfo{author}{\bibfnamefont{P.~A.} \bibnamefont{Midgley}} \bibnamefont{and}
  \bibinfo{author}{\bibfnamefont{R.~E.} \bibnamefont{Dunin-Borkowski}},
  \bibinfo{journal}{Nature materials} \textbf{\bibinfo{volume}{8}},
  \bibinfo{pages}{271} (\bibinfo{year}{2009}).

\bibitem[{\citenamefont{Matteucci et~al.}(2002)\citenamefont{Matteucci,
  Missiroli, and Pozzi}}]{Matteucci(2002)}
\bibinfo{author}{\bibfnamefont{G.}~\bibnamefont{Matteucci}},
  \bibinfo{author}{\bibfnamefont{G.}~\bibnamefont{Missiroli}},
  \bibnamefont{and} \bibinfo{author}{\bibfnamefont{G.}~\bibnamefont{Pozzi}}, in
  \emph{\bibinfo{booktitle}{Electron Microscopy and Holography II}}, edited by
  \bibinfo{editor}{\bibfnamefont{P.~W.} \bibnamefont{Hawkes}}
  (\bibinfo{publisher}{Elsevier}, \bibinfo{year}{2002}), vol.
  \bibinfo{volume}{Volume 122}, pp. \bibinfo{pages}{173--249}.

\bibitem[{\citenamefont{Natterer}(2001)}]{Natterer(2001)}
\bibinfo{author}{\bibfnamefont{F.}~\bibnamefont{Natterer}},
  \emph{\bibinfo{title}{The mathematics of computerized tomography}},
  vol.~\bibinfo{volume}{32} of \emph{\bibinfo{series}{Classics in applied
  mathematics}} (\bibinfo{publisher}{Society for Industrial and Applied
  Mathematics}, \bibinfo{address}{Philadelphia}, \bibinfo{year}{2001}), ISBN
  \bibinfo{isbn}{0898714931}.

\bibitem[{\citenamefont{Lai et~al.}(1994)\citenamefont{Lai, Hirayama, Ishizuka,
  and Tonomura}}]{Lai(1994)a}
\bibinfo{author}{\bibfnamefont{G.}~\bibnamefont{Lai}},
  \bibinfo{author}{\bibfnamefont{T.}~\bibnamefont{Hirayama}},
  \bibinfo{author}{\bibfnamefont{K.}~\bibnamefont{Ishizuka}}, \bibnamefont{and}
  \bibinfo{author}{\bibfnamefont{A.}~\bibnamefont{Tonomura}},
  \bibinfo{journal}{Journal of Applied Optics} \textbf{\bibinfo{volume}{33}},
  \bibinfo{pages}{829} (\bibinfo{year}{1994}).

\bibitem[{\citenamefont{Twitchett-Harrison
  et~al.}(2007)\citenamefont{Twitchett-Harrison, Yates, Newcomb,
  Dunin-Borkowski, and Midgley}}]{Twitchett-Harrison_NL7(2007)2020}
\bibinfo{author}{\bibfnamefont{A.~C.} \bibnamefont{Twitchett-Harrison}},
  \bibinfo{author}{\bibfnamefont{T.~J.~V.} \bibnamefont{Yates}},
  \bibinfo{author}{\bibfnamefont{S.~B.} \bibnamefont{Newcomb}},
  \bibinfo{author}{\bibfnamefont{R.~E.} \bibnamefont{Dunin-Borkowski}},
  \bibnamefont{and} \bibinfo{author}{\bibfnamefont{P.~A.}
  \bibnamefont{Midgley}}, \bibinfo{journal}{Nano Letters}
  \textbf{\bibinfo{volume}{7}}, \bibinfo{pages}{2020} (\bibinfo{year}{2007}).

\bibitem[{\citenamefont{Fujita and Chen}(2009)}]{Fujita_JoEM58(2009)301}
\bibinfo{author}{\bibfnamefont{T.}~\bibnamefont{Fujita}} \bibnamefont{and}
  \bibinfo{author}{\bibfnamefont{M.}~\bibnamefont{Chen}},
  \bibinfo{journal}{Journal of Electron Microscopy}
  \textbf{\bibinfo{volume}{58}}, \bibinfo{pages}{301} (\bibinfo{year}{2009}).

\bibitem[{\citenamefont{Wolf et~al.}(2010)\citenamefont{Wolf, Lubk, Lichte, and
  Friedrich}}]{Wolf_UM110(2010)390}
\bibinfo{author}{\bibfnamefont{D.}~\bibnamefont{Wolf}},
  \bibinfo{author}{\bibfnamefont{A.}~\bibnamefont{Lubk}},
  \bibinfo{author}{\bibfnamefont{H.}~\bibnamefont{Lichte}}, \bibnamefont{and}
  \bibinfo{author}{\bibfnamefont{H.}~\bibnamefont{Friedrich}},
  \bibinfo{journal}{Ultramicroscopy} \textbf{\bibinfo{volume}{110}},
  \bibinfo{pages}{390} (\bibinfo{year}{2010}).

\bibitem[{\citenamefont{Midgley and Weyland}(2003)}]{Midgley_UM96(2003)413}
\bibinfo{author}{\bibfnamefont{P.~A.} \bibnamefont{Midgley}} \bibnamefont{and}
  \bibinfo{author}{\bibfnamefont{M.}~\bibnamefont{Weyland}},
  \bibinfo{journal}{Ultramicroscopy} \textbf{\bibinfo{volume}{96}},
  \bibinfo{pages}{413} (\bibinfo{year}{2003}).

\bibitem[{\citenamefont{Tikhonov and Arsenin}(1977)}]{Tikhonov(1977)}
\bibinfo{author}{\bibfnamefont{A.~N.} \bibnamefont{Tikhonov}} \bibnamefont{and}
  \bibinfo{author}{\bibfnamefont{V.~Y.} \bibnamefont{Arsenin}},
  \emph{\bibinfo{title}{Solution of Ill-posed Problems.}}
  (\bibinfo{publisher}{Washington: Winston \& Sons}, \bibinfo{year}{1977}).

\bibitem[{\citenamefont{Goris et~al.}(2012)\citenamefont{Goris, Van~den Broek,
  Batenburg, Heidari~Mezerji, and Bals}}]{Goris(2012)}
\bibinfo{author}{\bibfnamefont{B.}~\bibnamefont{Goris}},
  \bibinfo{author}{\bibfnamefont{W.}~\bibnamefont{Van~den Broek}},
  \bibinfo{author}{\bibfnamefont{K.}~\bibnamefont{Batenburg}},
  \bibinfo{author}{\bibfnamefont{H.}~\bibnamefont{Heidari~Mezerji}},
  \bibnamefont{and} \bibinfo{author}{\bibfnamefont{S.}~\bibnamefont{Bals}},
  \bibinfo{journal}{Ultramicroscopy} \textbf{\bibinfo{volume}{113}},
  \bibinfo{pages}{120} (\bibinfo{year}{2012}).

\bibitem[{\citenamefont{Batenburg et~al.}(2009)\citenamefont{Batenburg, Bals,
  Sijbers, K\"ubel, Midgley, Hernandez, Kaiser, Encina, Coronado, and
  Tendeloo}}]{Batenburg(2009)}
\bibinfo{author}{\bibfnamefont{K.}~\bibnamefont{Batenburg}},
  \bibinfo{author}{\bibfnamefont{S.}~\bibnamefont{Bals}},
  \bibinfo{author}{\bibfnamefont{J.}~\bibnamefont{Sijbers}},
  \bibinfo{author}{\bibfnamefont{C.}~\bibnamefont{K\"ubel}},
  \bibinfo{author}{\bibfnamefont{P.}~\bibnamefont{Midgley}},
  \bibinfo{author}{\bibfnamefont{J.}~\bibnamefont{Hernandez}},
  \bibinfo{author}{\bibfnamefont{U.}~\bibnamefont{Kaiser}},
  \bibinfo{author}{\bibfnamefont{E.}~\bibnamefont{Encina}},
  \bibinfo{author}{\bibfnamefont{E.}~\bibnamefont{Coronado}}, \bibnamefont{and}
  \bibinfo{author}{\bibfnamefont{G.~V.} \bibnamefont{Tendeloo}},
  \bibinfo{journal}{Ultramicroscopy} \textbf{\bibinfo{volume}{109}},
  \bibinfo{pages}{730 } (\bibinfo{year}{2009}).

\bibitem[{\citenamefont{Hansen}(1998)}]{Hansen(1998)}
\bibinfo{author}{\bibfnamefont{P.~C.} \bibnamefont{Hansen}},
  \emph{\bibinfo{title}{Rank-Deficient and Discrete Ill-Posed Problems:
  Numerical Aspects of Linear Inversion}} (\bibinfo{publisher}{SIAM,
  Philadelphia,}, \bibinfo{year}{1998}).

\bibitem[{\citenamefont{Prete et~al.}(2008)\citenamefont{Prete, Marzo, Paiano,
  Lovergine, Salviati, Lazzarini, and Sekiguchi}}]{Prete_JoCG310(2008)5114}
\bibinfo{author}{\bibfnamefont{P.}~\bibnamefont{Prete}},
  \bibinfo{author}{\bibfnamefont{F.}~\bibnamefont{Marzo}},
  \bibinfo{author}{\bibfnamefont{P.}~\bibnamefont{Paiano}},
  \bibinfo{author}{\bibfnamefont{N.}~\bibnamefont{Lovergine}},
  \bibinfo{author}{\bibfnamefont{G.}~\bibnamefont{Salviati}},
  \bibinfo{author}{\bibfnamefont{L.}~\bibnamefont{Lazzarini}},
  \bibnamefont{and}
  \bibinfo{author}{\bibfnamefont{T.}~\bibnamefont{Sekiguchi}},
  \bibinfo{journal}{Journal of Crystal Growth} \textbf{\bibinfo{volume}{310}},
  \bibinfo{pages}{5114} (\bibinfo{year}{2008}).

\bibitem[{\citenamefont{Wolf et~al.}(2011)\citenamefont{Wolf, Lichte, Pozzi,
  Prete, and Lovergine}}]{Wolf(2011)}
\bibinfo{author}{\bibfnamefont{D.}~\bibnamefont{Wolf}},
  \bibinfo{author}{\bibfnamefont{H.}~\bibnamefont{Lichte}},
  \bibinfo{author}{\bibfnamefont{G.}~\bibnamefont{Pozzi}},
  \bibinfo{author}{\bibfnamefont{P.}~\bibnamefont{Prete}}, \bibnamefont{and}
  \bibinfo{author}{\bibfnamefont{N.}~\bibnamefont{Lovergine}},
  \bibinfo{journal}{Applied Physics Letters} \textbf{\bibinfo{volume}{98}},
  \bibinfo{eid}{264103} (pages~\bibinfo{numpages}{3}) (\bibinfo{year}{2011}).

\bibitem[{\citenamefont{Rudolph et~al.}(2013)\citenamefont{Rudolph, Funk,
  Doblinger, Mork\"otter, Hertenberger, Schweickert, Becker, Matich, Bichler,
  Spirkoska et~al.}}]{Rudolph(2013)}
\bibinfo{author}{\bibfnamefont{D.}~\bibnamefont{Rudolph}},
  \bibinfo{author}{\bibfnamefont{S.}~\bibnamefont{Funk}},
  \bibinfo{author}{\bibfnamefont{M.}~\bibnamefont{Doblinger}},
  \bibinfo{author}{\bibfnamefont{S.}~\bibnamefont{Mork\"otter}},
  \bibinfo{author}{\bibfnamefont{S.}~\bibnamefont{Hertenberger}},
  \bibinfo{author}{\bibfnamefont{L.}~\bibnamefont{Schweickert}},
  \bibinfo{author}{\bibfnamefont{J.}~\bibnamefont{Becker}},
  \bibinfo{author}{\bibfnamefont{S.}~\bibnamefont{Matich}},
  \bibinfo{author}{\bibfnamefont{M.}~\bibnamefont{Bichler}},
  \bibinfo{author}{\bibfnamefont{D.}~\bibnamefont{Spirkoska}},
  \bibnamefont{et~al.}, \bibinfo{journal}{Nano Letters}
  \textbf{\bibinfo{volume}{13}}, \bibinfo{pages}{1522} (\bibinfo{year}{2013}).

\bibitem[{\citenamefont{Heiss et~al.}(2013)\citenamefont{Heiss, Fontana,
  Gustafsson, W\"ust, Magen, O'Regan, Luo, Ketterer, Conesa-Boj, Kuhlmann
  et~al.}}]{Heiss_NM12(2013)439}
\bibinfo{author}{\bibfnamefont{M.}~\bibnamefont{Heiss}},
  \bibinfo{author}{\bibfnamefont{Y.}~\bibnamefont{Fontana}},
  \bibinfo{author}{\bibfnamefont{A.}~\bibnamefont{Gustafsson}},
  \bibinfo{author}{\bibfnamefont{G.}~\bibnamefont{W\"ust}},
  \bibinfo{author}{\bibfnamefont{C.}~\bibnamefont{Magen}},
  \bibinfo{author}{\bibfnamefont{D.~D.} \bibnamefont{O'Regan}},
  \bibinfo{author}{\bibfnamefont{J.~W.} \bibnamefont{Luo}},
  \bibinfo{author}{\bibfnamefont{B.}~\bibnamefont{Ketterer}},
  \bibinfo{author}{\bibfnamefont{S.}~\bibnamefont{Conesa-Boj}},
  \bibinfo{author}{\bibfnamefont{A.~V.} \bibnamefont{Kuhlmann}},
  \bibnamefont{et~al.}, \bibinfo{journal}{Nature materials}
  \textbf{\bibinfo{volume}{12}}, \bibinfo{pages}{439} (\bibinfo{year}{2013}).

\bibitem[{\citenamefont{Bryllert et~al.}(2006)\citenamefont{Bryllert,
  Wernersson, L\"owgren, and Samuelson}}]{Bryllert_NT17(2006)227}
\bibinfo{author}{\bibfnamefont{T.}~\bibnamefont{Bryllert}},
  \bibinfo{author}{\bibfnamefont{L.-E.} \bibnamefont{Wernersson}},
  \bibinfo{author}{\bibfnamefont{T.}~\bibnamefont{L\"owgren}},
  \bibnamefont{and}
  \bibinfo{author}{\bibfnamefont{L.}~\bibnamefont{Samuelson}},
  \bibinfo{journal}{Nanotechnology} \textbf{\bibinfo{volume}{17}},
  \bibinfo{pages}{S227} (\bibinfo{year}{2006}).

\bibitem[{\citenamefont{Gallo et~al.}(2011)\citenamefont{Gallo, Chen, Currie,
  McGuckin, Prete, Lovergine, Nabet, and Spanier}}]{Gallo_APL98(2011)241113}
\bibinfo{author}{\bibfnamefont{E.~M.} \bibnamefont{Gallo}},
  \bibinfo{author}{\bibfnamefont{G.}~\bibnamefont{Chen}},
  \bibinfo{author}{\bibfnamefont{M.}~\bibnamefont{Currie}},
  \bibinfo{author}{\bibfnamefont{T.}~\bibnamefont{McGuckin}},
  \bibinfo{author}{\bibfnamefont{P.}~\bibnamefont{Prete}},
  \bibinfo{author}{\bibfnamefont{N.}~\bibnamefont{Lovergine}},
  \bibinfo{author}{\bibfnamefont{B.}~\bibnamefont{Nabet}}, \bibnamefont{and}
  \bibinfo{author}{\bibfnamefont{J.~E.} \bibnamefont{Spanier}},
  \bibinfo{journal}{Applied Physics Letters} \textbf{\bibinfo{volume}{98}},
  \bibinfo{pages}{241113} (\bibinfo{year}{2011}).

\bibitem[{\citenamefont{Krogstrup et~al.}(2013)\citenamefont{Krogstrup,
  Jorgensen, Heiss, Demichel, Holm, Aagesen, Nygard, and Fontcuberta~i
  Morral}}]{Krogstrup_NP7(2013)306}
\bibinfo{author}{\bibfnamefont{P.}~\bibnamefont{Krogstrup}},
  \bibinfo{author}{\bibfnamefont{H.~I.} \bibnamefont{Jorgensen}},
  \bibinfo{author}{\bibfnamefont{M.}~\bibnamefont{Heiss}},
  \bibinfo{author}{\bibfnamefont{O.}~\bibnamefont{Demichel}},
  \bibinfo{author}{\bibfnamefont{J.~V.} \bibnamefont{Holm}},
  \bibinfo{author}{\bibfnamefont{M.}~\bibnamefont{Aagesen}},
  \bibinfo{author}{\bibfnamefont{J.}~\bibnamefont{Nygard}}, \bibnamefont{and}
  \bibinfo{author}{\bibfnamefont{A.}~\bibnamefont{Fontcuberta~i Morral}},
  \bibinfo{journal}{Nat Photon} \textbf{\bibinfo{volume}{7}},
  \bibinfo{pages}{306} (\bibinfo{year}{2013}).

\bibitem[{\citenamefont{Lehmann and Lichte}(2002)}]{Lehmann_MaM8(2002)447}
\bibinfo{author}{\bibfnamefont{M.}~\bibnamefont{Lehmann}} \bibnamefont{and}
  \bibinfo{author}{\bibfnamefont{H.}~\bibnamefont{Lichte}},
  \bibinfo{journal}{Microscopy and Microanalysis} \textbf{\bibinfo{volume}{8}},
  \bibinfo{pages}{447} (\bibinfo{year}{2002}).

\bibitem[{\citenamefont{Ghiglia and Pritt}(1998)}]{Ghiglia(1998)}
\bibinfo{author}{\bibfnamefont{D.~C.} \bibnamefont{Ghiglia}} \bibnamefont{and}
  \bibinfo{author}{\bibfnamefont{M.~D.} \bibnamefont{Pritt}},
  \emph{\bibinfo{title}{Two-Dimensional Phase Unwrapping}}
  (\bibinfo{publisher}{Wiley}, \bibinfo{year}{1998}).

\bibitem[{\citenamefont{Lubk et~al.}(2010)\citenamefont{Lubk, Wolf, and
  Lichte}}]{Lubk(2010)}
\bibinfo{author}{\bibfnamefont{A.}~\bibnamefont{Lubk}},
  \bibinfo{author}{\bibfnamefont{D.}~\bibnamefont{Wolf}}, \bibnamefont{and}
  \bibinfo{author}{\bibfnamefont{H.}~\bibnamefont{Lichte}},
  \bibinfo{journal}{Ultramicroscopy} \textbf{\bibinfo{volume}{110}},
  \bibinfo{pages}{438 } (\bibinfo{year}{2010}).

\bibitem[{\citenamefont{Gilbert}(1972)}]{Gilbert_PRSL.182(1972)89}
\bibinfo{author}{\bibfnamefont{P.}~\bibnamefont{Gilbert}},
  \bibinfo{journal}{Proc. R. Soc. Lond. B.} \textbf{\bibinfo{volume}{182}},
  \bibinfo{pages}{89} (\bibinfo{year}{1972}).

\bibitem[{\citenamefont{Radermacher}(2006)}]{Radermacher(2006)}
\bibinfo{author}{\bibfnamefont{M.}~\bibnamefont{Radermacher}}, in
  \emph{\bibinfo{booktitle}{Electron Tomography, Methods for Three-Dimensional
  Visualization of Structures in the Cell}}, edited by
  \bibinfo{editor}{\bibfnamefont{J.}~\bibnamefont{Frank}}
  (\bibinfo{publisher}{Springer}, \bibinfo{address}{Berlin},
  \bibinfo{year}{2006}), pp. \bibinfo{pages}{245--274}.

\bibitem[{\citenamefont{Wolf et~al.}(2014)\citenamefont{Wolf, Lubk, and
  Lichte}}]{Wolf(2014)}
\bibinfo{author}{\bibfnamefont{D.}~\bibnamefont{Wolf}},
  \bibinfo{author}{\bibfnamefont{A.}~\bibnamefont{Lubk}}, \bibnamefont{and}
  \bibinfo{author}{\bibfnamefont{H.}~\bibnamefont{Lichte}},
  \bibinfo{journal}{Ultramicroscopy} \textbf{\bibinfo{volume}{136}},
  \bibinfo{pages}{15} (\bibinfo{year}{2014}).

\bibitem[{\citenamefont{Grillo and Rossi}(2013)}]{Grillo(2013)}
\bibinfo{author}{\bibfnamefont{V.}~\bibnamefont{Grillo}} \bibnamefont{and}
  \bibinfo{author}{\bibfnamefont{F.}~\bibnamefont{Rossi}},
  \bibinfo{journal}{Ultramicroscopy} \textbf{\bibinfo{volume}{125}},
  \bibinfo{pages}{112} (\bibinfo{year}{2013}).

\bibitem[{\citenamefont{Paige and Saunders}(1982)}]{Paige(1982)}
\bibinfo{author}{\bibfnamefont{C.~C.} \bibnamefont{Paige}} \bibnamefont{and}
  \bibinfo{author}{\bibfnamefont{M.~A.} \bibnamefont{Saunders}},
  \bibinfo{journal}{{ACM} Transactions on Mathematical Software}
  \textbf{\bibinfo{volume}{8}}, \bibinfo{pages}{195} (\bibinfo{year}{1982}).

\bibitem[{\citenamefont{{Schowalter} et~al.}(2006)\citenamefont{{Schowalter},
  {Rosenauer}, {Lamoen}, {Kruse}, and {Gerthsen}}}]{Schowalter(2006)}
\bibinfo{author}{\bibfnamefont{M.}~\bibnamefont{{Schowalter}}},
  \bibinfo{author}{\bibfnamefont{A.}~\bibnamefont{{Rosenauer}}},
  \bibinfo{author}{\bibfnamefont{D.}~\bibnamefont{{Lamoen}}},
  \bibinfo{author}{\bibfnamefont{P.}~\bibnamefont{{Kruse}}}, \bibnamefont{and}
  \bibinfo{author}{\bibfnamefont{D.}~\bibnamefont{{Gerthsen}}},
  \bibinfo{journal}{Applied Physics Letters} \textbf{\bibinfo{volume}{88}},
  \bibinfo{pages}{232108} (\bibinfo{year}{2006}).

\bibitem[{\citenamefont{Kruse et~al.}(2006)\citenamefont{Kruse, Schowalter,
  Lamoen, Rosenauer, and Gerthsen}}]{Kruse(2006)}
\bibinfo{author}{\bibfnamefont{P.}~\bibnamefont{Kruse}},
  \bibinfo{author}{\bibfnamefont{M.}~\bibnamefont{Schowalter}},
  \bibinfo{author}{\bibfnamefont{D.}~\bibnamefont{Lamoen}},
  \bibinfo{author}{\bibfnamefont{A.}~\bibnamefont{Rosenauer}},
  \bibnamefont{and} \bibinfo{author}{\bibfnamefont{D.}~\bibnamefont{Gerthsen}},
  \bibinfo{journal}{Ultramicroscopy} \textbf{\bibinfo{volume}{106}},
  \bibinfo{pages}{105} (\bibinfo{year}{2006}).

\bibitem[{\citenamefont{Sk\"old et~al.}(2005)\citenamefont{Sk\"old, Karlsson,
  Larsson, Pistol, Seifert, Tr\"agardh, and Samuelson}}]{Skold(2005)}
\bibinfo{author}{\bibfnamefont{N.}~\bibnamefont{Sk\"old}},
  \bibinfo{author}{\bibfnamefont{L.~S.} \bibnamefont{Karlsson}},
  \bibinfo{author}{\bibfnamefont{M.~W.} \bibnamefont{Larsson}},
  \bibinfo{author}{\bibfnamefont{M.-E.} \bibnamefont{Pistol}},
  \bibinfo{author}{\bibfnamefont{W.}~\bibnamefont{Seifert}},
  \bibinfo{author}{\bibfnamefont{J.}~\bibnamefont{Tr\"agardh}},
  \bibnamefont{and}
  \bibinfo{author}{\bibfnamefont{L.}~\bibnamefont{Samuelson}},
  \bibinfo{journal}{Nano Letters} \textbf{\bibinfo{volume}{5}},
  \bibinfo{pages}{1943} (\bibinfo{year}{2005}).

\bibitem[{\citenamefont{Zheng et~al.}(2013)\citenamefont{Zheng, Wong-Leung,
  Gao, Tan, Jagadish, and Etheridge}}]{Zheng(2013)}
\bibinfo{author}{\bibfnamefont{C.}~\bibnamefont{Zheng}},
  \bibinfo{author}{\bibfnamefont{J.}~\bibnamefont{Wong-Leung}},
  \bibinfo{author}{\bibfnamefont{Q.}~\bibnamefont{Gao}},
  \bibinfo{author}{\bibfnamefont{H.~H.} \bibnamefont{Tan}},
  \bibinfo{author}{\bibfnamefont{C.}~\bibnamefont{Jagadish}}, \bibnamefont{and}
  \bibinfo{author}{\bibfnamefont{J.}~\bibnamefont{Etheridge}},
  \bibinfo{journal}{Nano Lett.} \textbf{\bibinfo{volume}{13}},
  \bibinfo{pages}{3742} (\bibinfo{year}{2013}).

\bibitem[{\citenamefont{Johansson et~al.}(2007)\citenamefont{Johansson,
  Karlsson, Svensson, M{\aa}rtensson, Wacaser, Deppert, Samuelson, and
  Seifert}}]{Johansson2007}
\bibinfo{author}{\bibfnamefont{J.}~\bibnamefont{Johansson}},
  \bibinfo{author}{\bibfnamefont{L.~S.} \bibnamefont{Karlsson}},
  \bibinfo{author}{\bibfnamefont{C.~P.~T.} \bibnamefont{Svensson}},
  \bibinfo{author}{\bibfnamefont{T.}~\bibnamefont{M{\aa}rtensson}},
  \bibinfo{author}{\bibfnamefont{B.~A.} \bibnamefont{Wacaser}},
  \bibinfo{author}{\bibfnamefont{K.}~\bibnamefont{Deppert}},
  \bibinfo{author}{\bibfnamefont{L.}~\bibnamefont{Samuelson}},
  \bibnamefont{and} \bibinfo{author}{\bibfnamefont{W.}~\bibnamefont{Seifert}},
  \bibinfo{journal}{Journal of Crystal Growth} \textbf{\bibinfo{volume}{298}},
  \bibinfo{pages}{635} (\bibinfo{year}{2007}).

\end{thebibliography}

\end{document}